\documentclass{iau}
\usepackage{graphicx}

\title[Solar wind on pulsar observations at low frequency] 
{The effect of the Solar wind on low-frequency observations of pulsars}

\author[C. Tiburzi \& J. Verbiest]   
{C. Tiburzi$^{1,2}$,
 J.~P.~W.~Verbiest$^{2,1}$}

\affiliation{$^1$Max-Planck-Institut f\"{u}r Radioastronomie, \\ Auf dem H\"{u}gel 69,
53121 Bonn, Germany \\ email: {\tt ctiburzi@mpifr-bonn.mpg.de} \\[\affilskip]
$^2$Universit\"{a}t Bielefeld, Fakult\"{a}t f\"{u}r Physik, \\ Universit\"{a}tsstr. 25, D-33615 Bielefeld, Germany}

\pubyear{2017}
\volume{337}  
\jname{Pulsar Astrophysics -­ The Next 50 Years}
\editors{P. Weltevrede, B.B.P. Perera, L. Levin Preston \& S. Sanidas, eds.}

\begin{document}

\maketitle

\begin{abstract} 
We operate the six German stations of the LOw Frequency ARray as standalone telescopes to observe more than 100 pulsars every week. To date, we have collected almost four years of high-quality data at an unprecedented weekly cadence. This allows us to perform a wide variety of analyses aimed at characterising the magnetoionic plasma crossed by pulsar radiation. In particular, our studies are focused on electron density variations in the interstellar and interplanetary media, the Galactic and interplanetary magnetic field, scintillation, and extreme scattering events. Here we report the first results from our Solar wind monitoring campaign.
\keywords{Pulsars: general, Solar wind}
\end{abstract}

\firstsection 
\section{Introduction}

The LOw Frequency ARray (LOFAR, \cite[van Haarlem \etal\ 2013]{vwg13}) is a low-frequency interferometer operating between 10 and 240 MHz. The modules of the interferometer, called \textit{stations}, are located in the Netherlands (38 stations), Germany (6), Poland (3), France, the U.K., Sweden and Ireland (one station each), and are themselves interferometers. In the Netherlands, 24 stations are located close to Exloo and constitute the \textit{core} of LOFAR. An additional 14 stations (referred to as the \textit{remote} stations) are distributed across the Netherlands. The stations present in the other contributing nations, referred to as the \textit{international} stations, can either participate as additional long baselines of LOFAR, or operated as \textit{independent and stand-alone} telescopes. 
Each station consists of two types of antennas -- the \textit{low band antennas} (LBAs), operating between $\sim$10 and $\sim$90 MHz, and the \textit{high band antennas} (HBAs), operating between $\sim$110 and $\sim$240 MHz. The Dutch stations (both the core and remote stations) are made of 96 LBAs and 48 \textit{tiles} (units comprised of 16 pairs of dipoles) of HBAs, while the international stations are made of 96 LBAs and HBA tiles.\\
 \indent Every week, we use the HBA fields of the six international German stations (the German LOng Wavelength, or GLOW, array) to observe more than 100 pulsars in a coherently-dedispersed (\cite[van Straten \& Bailes 2011]{vb11}) folding mode. This observing campaign commenced in 2013, with the first stations available in Germany, and to date has collected more than 52 TB of pulsar data. \\
\indent At LOFAR frequencies, pulsar observations enable a wealth of studies of the interstellar and interplanetary plasma. The impact of magnetoionic plasma on the propagation of broadband radio emission is much greater at low frequencies. Particularly interesting, and with multidisciplinary implications, is the study of the electron and magnetic content of the Solar wind.  It is worth to notice that pulsar-based studies are the only means to obtain model-independent estimates of the magnetic field of the Solar wind, other than \textit{in situ} measurements from spacecraft. To date, however, the majority of pulsar-based studies of the Solar wind used poorly sampled, high-frequency data (usually at 1400~MHz), and thus the precision is greatly improved with our high-cadence, low-frequency pulsar data set. \\ 
\indent Below we report the first results of our Solar wind analysis, consisting of the measures of the variable electron content along the line-of-sight (LOS) of three millisecond pulsars (MSPs) during their Solar approaches. The electron content is expressed through the \textit{dispersion measure} (DM) parameter, the integrated the free electron density along the LOS.

\section{Observations and Methods}

\begin{table}
 \caption{Characteristics of the analysed MSPs.}
  \label{msps}
  \begin{center} 
 {\scriptsize
  \begin{tabular}{ccccc}\hline 
{\bf Pulsar} & {\bf Period } & {\bf DM} & {\bf RM } & {\bf Ecliptic} \\ 
& {\bf [msec]} & {\bf  [pc/cm$^3$]} & {\bf  [rad/m$^2$]} & {\bf latitude [deg]} \\ 
 \hline
 J0034$-$0534 & 1.9 & 13.765 & $-$ & $-8.53$\\
 J1022+1001 & 16.5 & 10.252  & 1.39  & $-0.06$ \\
 J2145$-$0750 & 16.1 & 8.998 & $-0.8$ & 5.31\\
 \hline
  \end{tabular}
  }
 \end{center}
\end{table}

The three MSPs we have analysed are PSRs J0034$-$0534, J1022+1001 and J2145$-$0750. Although each German station uses independent observing lists, some pulsars are shared for the purpose of comparing station performance. The three analysed MSPs are among the shared sources, meaning that four datasets are available for each. The three MSPs were chosen because of their low ecliptic latitude. This means that, yearly, they closely approach the Sun, and during the approach the effects of the Solar wind are expected to be maximized. Some key characteristics of these MSPs are summarized in Table \ref{msps}.\\
\indent The dataset for the three MSPs spans four years from 2013 to the beginning of 2017. The observations were obtained with the four German LOFAR stations located at Effelsberg (DE601, fully operational since 2013), Unterweilenbach (DE602, operational since 2015), Tautenburg (DE603, operational since 2014) and Juelich (DE605, operational since 2014). For each station, the data were obtained at a weekly cadence with the HBAs. The central frequency of the observations is $\sim$150 MHz and for DE601 the bandwidth is $\sim$95 MHz divided into 488 coherently-dedispersed channels, while for the other stations the bandwidth is $\sim$70 MHz divided into 366 coherently-dedispersed channels (before 2015, the bandwidth for the other three stations was also $\sim$95 MHz). The integration time for each observation is typically three hours, folded into subintegrations of 10 seconds.\\
\indent For each observation, we removed radio frequency interference by using dedicated tools present in the \textsc{PSRCHIVE} (\cite[van Straten \etal\ 2012]{vdo12}) and \textsc{coastguard}  (\cite[Lazarus \etal\ 2016]{lkg16}) software packages. We then used the longest dataset for each pulsar (usually that obtained with the Effelsberg station) to improve the DM of the pulsar ephemeris obtained by the International Pulsar Timing Array (IPTA, \cite[Verbiest \etal\ 2016]{vlh16}) through the technique of \textit{pulsar timing} (see \cite[Lorimer \& Kramer 2005]{lk05}). We then installed the improved ephemeris in the individual observations of each dataset, calibrated the polarization, and averaged the observations in time.  \\
\indent These final versions of the datasets were used to obtain the DM time series. For this purpose, we first obtained a frequency-resolved template by averaging together all the observations of one dataset (i.e., of one pulsar-station combination). If necessary, we partially averaged the frequency channels to increase the signal-to-noise ratio, and applied the same averaging factor to the template. We then obtained the times-of-arrival (ToAs) of the pulsar emission per frequency channel in each observation independently, by cross-correlating the observation with the standard template. To avoid contamination from the off-pulse noise, we performed the cross-correlation using half of the available harmonics. We then used the software package \textsc{tempo2} (\cite[Hobbs \etal\ 2006]{hem06}) to obtain the DM from the frequency-resolved ToAs of each observation. \textsc{tempo2}  provides basic mitigation for the heliospheric effects, by modelling the Solar wind as a spherical distribution of free electrons with $n_0=4$ cm$^{-3}$ at one astronomical unit (AU) distance from the Sun. We obtained two time series of DM values: one while maintaining the Solar wind correction, the second while omitting it.

\section{Results and Discussion}

\begin{figure}
\begin{tabular}{ll}
 \includegraphics[width=2.8in]{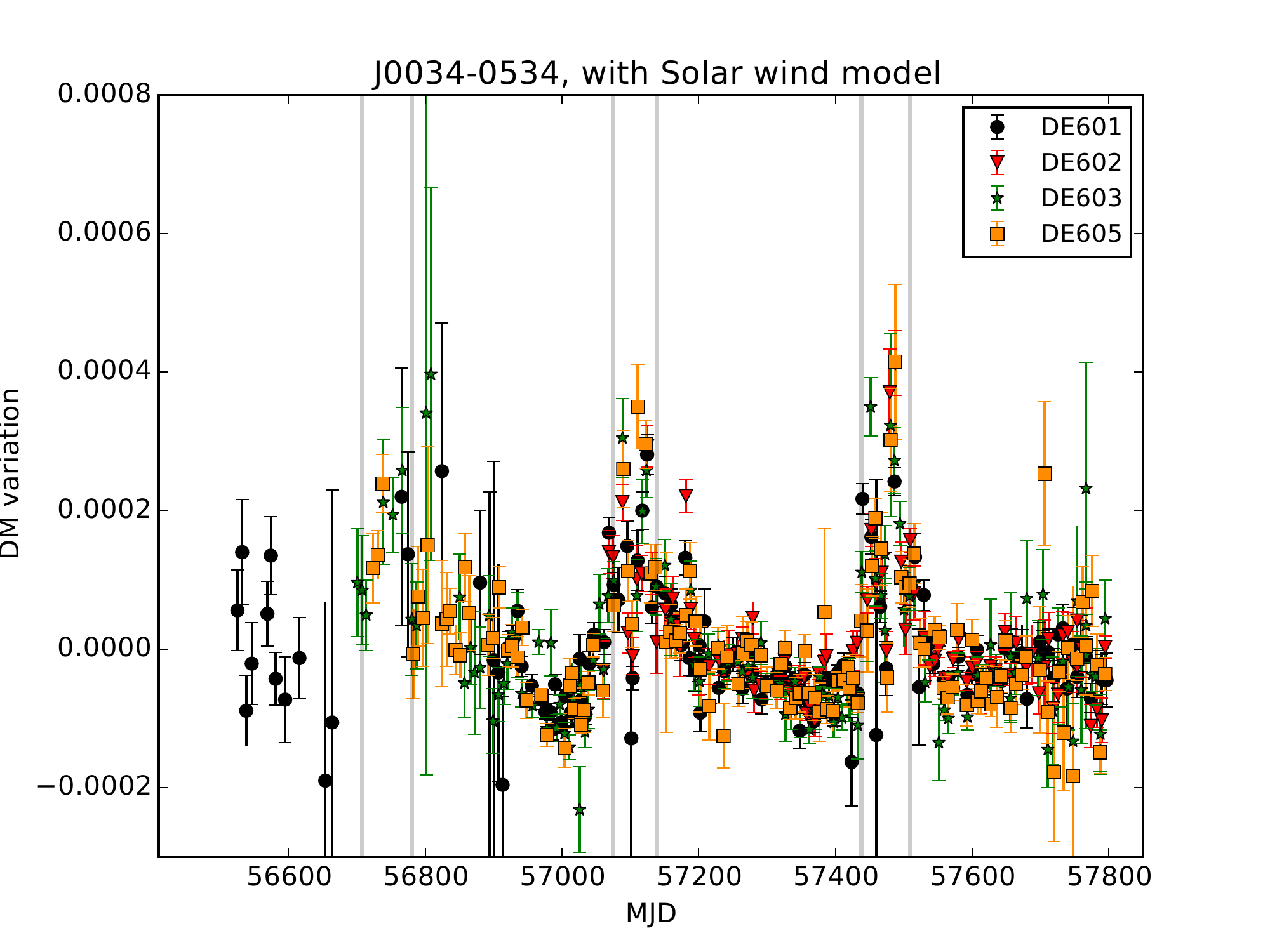} &
 \includegraphics[width=2.8in]{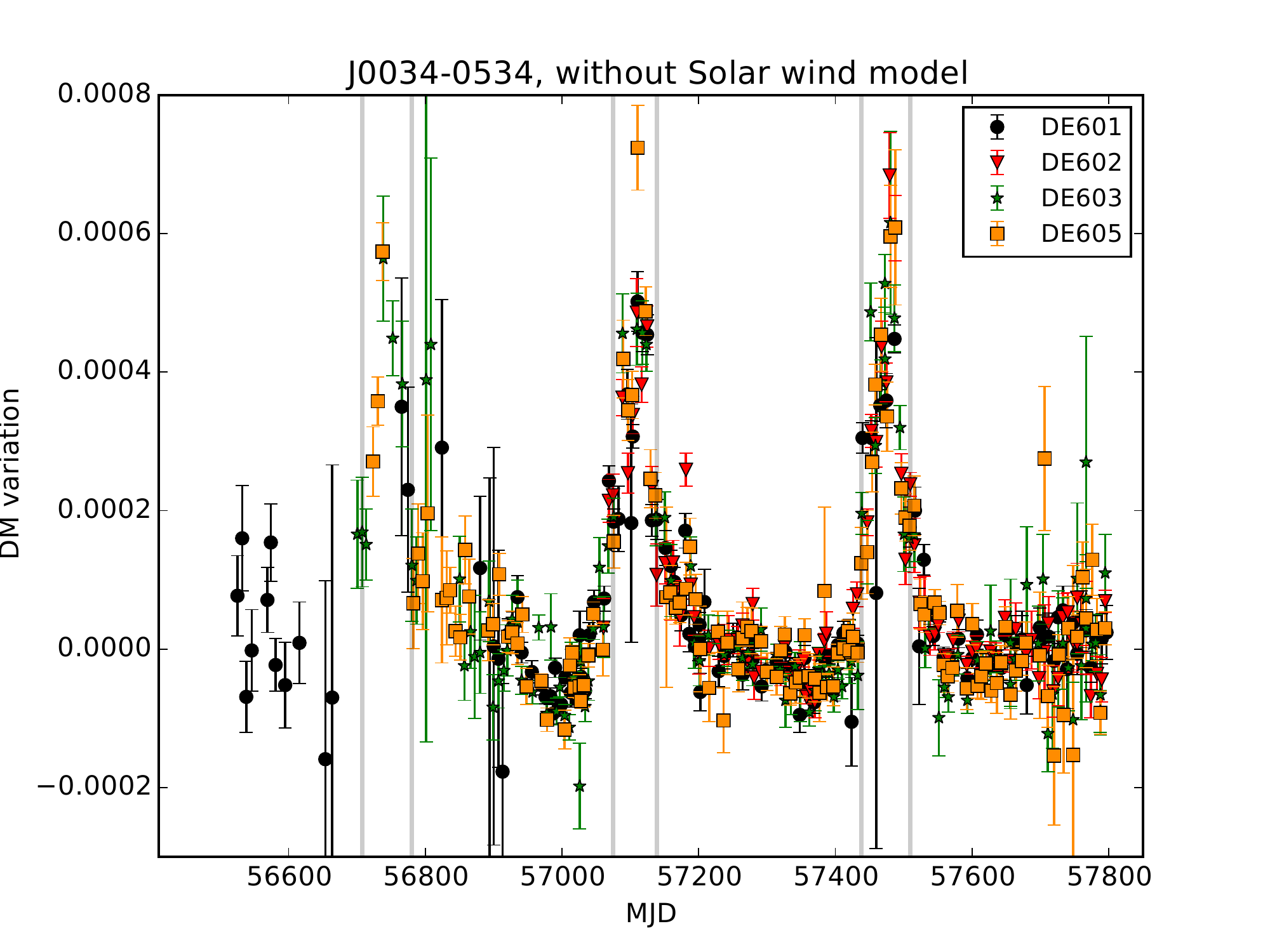}\\
 \includegraphics[width=2.8in]{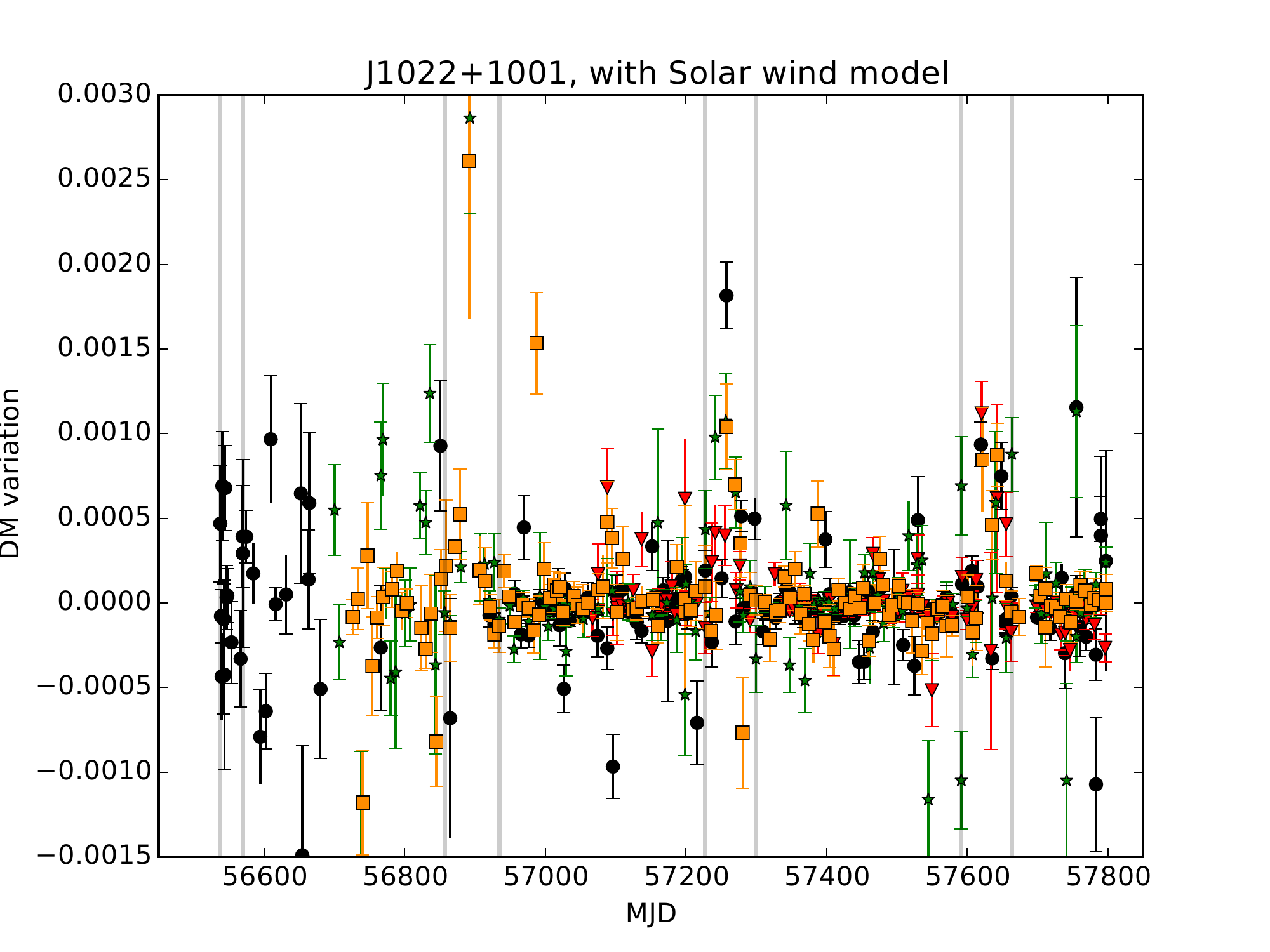} &
 \includegraphics[width=2.8in]{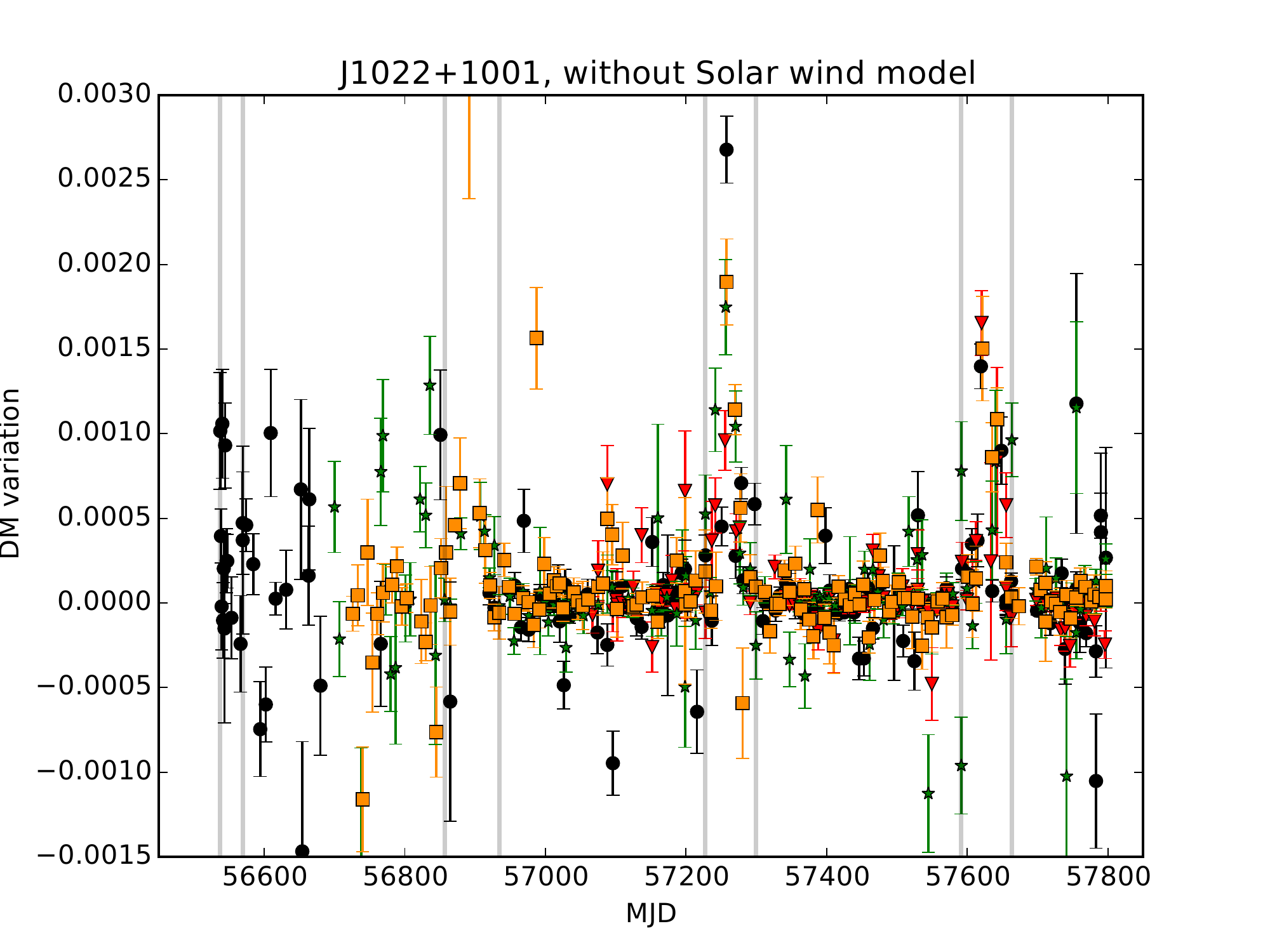}\\
 \includegraphics[width=2.8in]{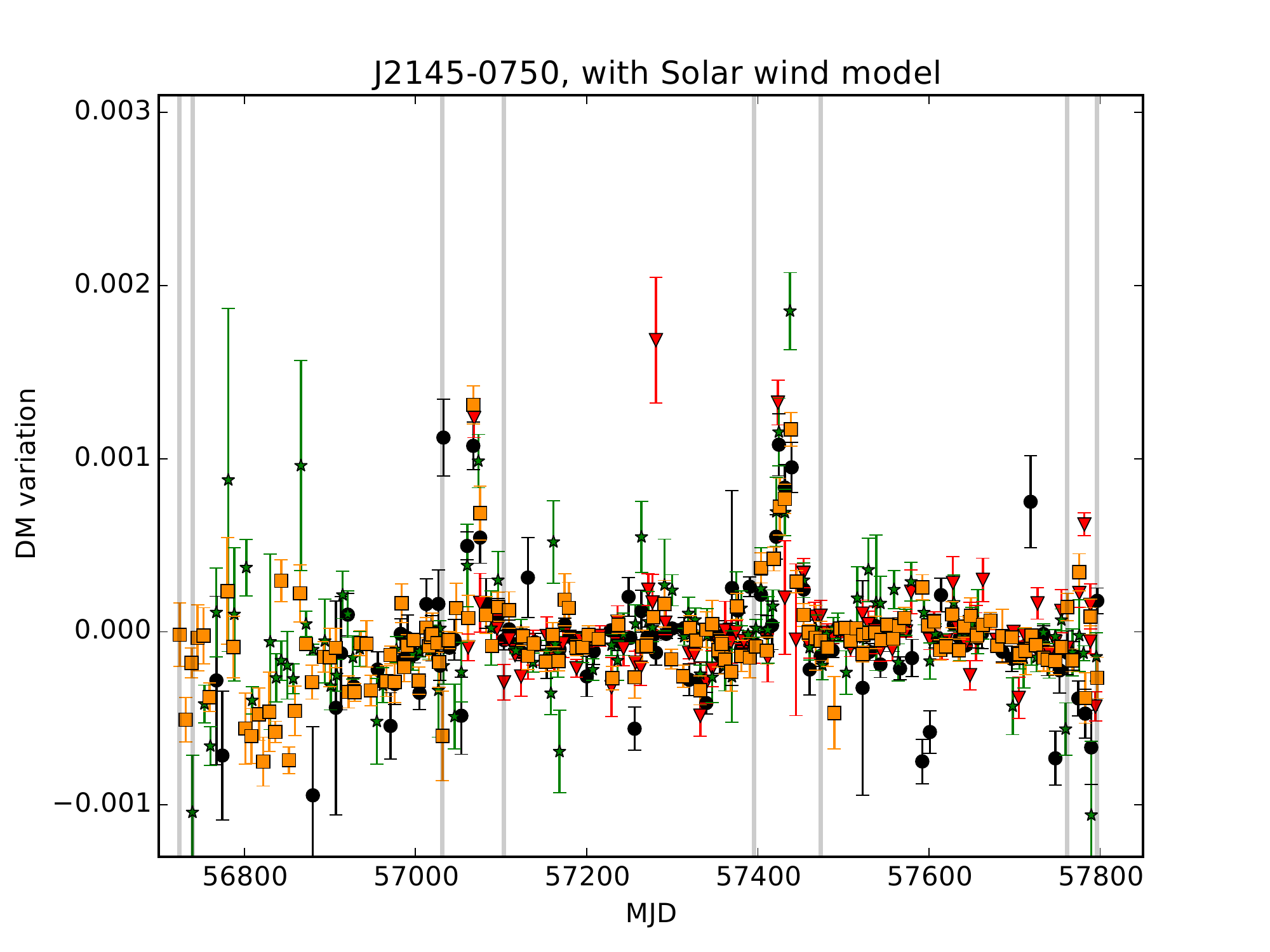} &
 \includegraphics[width=2.8in]{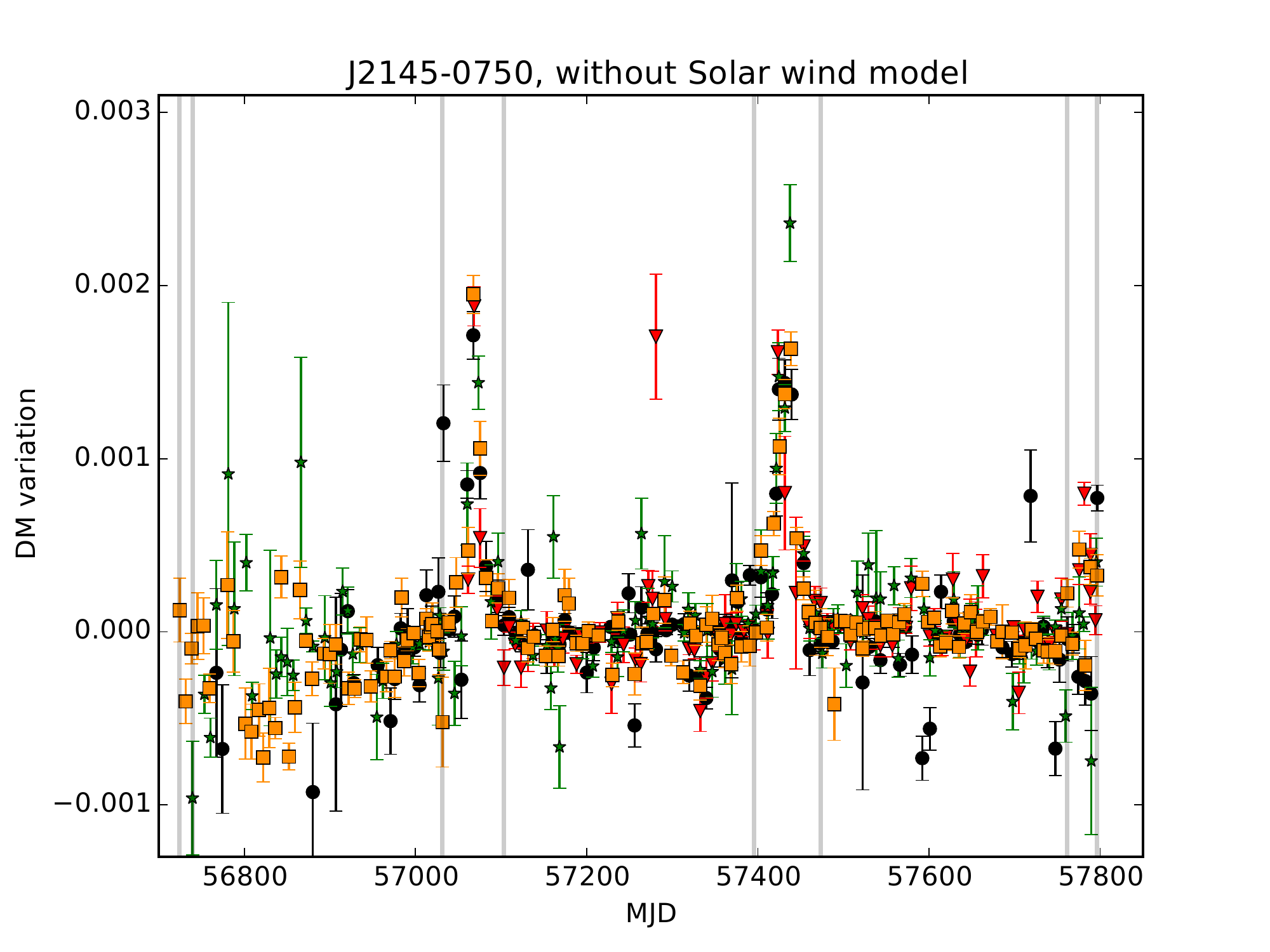}
 \end{tabular}
\caption{DM time series for PSRs~J0034$-$0534 (top row), J1022+1001 (central), J2145$-$0750 (bottom). On the left-hand side are shown the time series obtained with the Solar wind mitigation on, while on the right-hand side the DM series without the Solar wind mitigation are given. The colours and symbols of the data points refer to the different stations, while the vertical lines indicate an angular distance from the Sun of 40 degrees.}
   \label{results}
\end{figure}

The time series of the DM variations obtained for the three MSPs are shown in Fig.\,\ref{results}. It is immediately clear that the mitigation routine provided by \textsc{tempo2} does not adequately correct for the Solar wind effect. The Solar wind signatures are clearly visible in all plots on the left hand side of Fig.\,\ref{results}. It is not surprising that this effect has not been noticed so clearly at higher frequencies, as the sensitivity to DM variations is greater when observing at low frequencies. Our results demonstrate that the true effect of the Solar wind is about a factor of two higher than that estimated by the simple mitigation technique applied by \textsc{tempo2}. It is possible that a different $n_0$ value will improve the correction. The maximum of the Solar cycle was reached just in 2014, and the \textsc{tempo2} $n_0$ assumes the Solar minimum. \\
\indent A second consideration stems from the results obtained for PSR~J0034$-$0534. This MSP rotates at a very high frequency (one order of magnitude greater than the other two MSPs), and its timing precision is accordingly higher enabling an exquisite precision in the DM measurements. The amplitude of DM variations in this pulsar is approximately half that of the other two pulsars, due to its high ecliptic latitude, but it is much more clearly visible. It is interesting to note that the DM variations induced by the Solar wind in PSR~J0034$-$0534 are asymmetric, with a rapid and steep rise while the pulsar begins to approach the Sun, and a smoother and slower decrease while it recedes from it. While for the other two pulsars it might be possible to tune the density of free electrons at one AU to mitigate the Solar wind effect, for this pulsar the spherical model cannot correct for the Solar wind effect. By increasing the sensitivity of our instruments at low frequencies, we will be able to study these kind of asymmetries in detail, and to propose more sophisticated models to correct for the Solar wind effect.

\small{
\section*{Acknowledgements}
This analysis makes use of data from the LOFAR stations at Effelsberg and Unterweilenbach funded by the Max-Planck-Gesellschaft; at Tautenburg funded by the BMBF Verbundforschung project D-LOFAR I and the European Union (EFRE); and at Juelich supported by the BMBF Verbundforschung project D-LOFAR I. The observations of the German LOFAR stations were carried out in the stand-alone GLOW mode (German LOng-Wavelength array) which is technically operated and supported by the Max-Planck-Institut fuer Radioastronomie, the Forschungszentrum Juelich and Bielefeld University. The observations were possible thanks to A. Horneffer, J. Kuensemoeller, J. Verbiest, C. Tiburzi, S. Os\l{}owski, N. Porayko, M. Serylak and J. Griessmeier. This analysis is carried out in collaboration with G. Janssen, W. Coles, G. Shaifullah, and R. Fallows.
}


\begin{thebibliography}{}  
 
 
\bibitem[Hobbs \etal\ (2006)]{hem06}
{Hobbs G. B. et al.} 2006
\textit{MNRAS}, 369, p. 655-672
 
\bibitem[Lazarus \etal\ (2016)]{lkg16}
{Lazarus P. et al.} 2016,
\textit{MNRAS}, 458, p. 868-880

\bibitem[Lorimer \& Kramer (2005)]{lk05}
{Lorimer D. R. \& Kramer M.} 2005
\textit{Cambridge University Press}
 
\bibitem[van Haarlem \etal\ (2013)]{vwg13}
{van Haarlem M.~P. et al.} 2013,
\textit{A\&A}, 55, A2

\bibitem[van Straten \& Bailes (2011)]{vb11}
{van Straten W. \& Bailes M.} 2011,
\textit{PASA}, 28, p. 1-14

\bibitem[van Straten \etal\ (2012)]{vdo12}
{van Straten W. et al.} 2012,
\textit{PASA}, 9, p. 237-259

\bibitem[Verbiest \etal\ (2016)]{vlh16}
{Verbiest J. P. W. et al.} 2016,
\textit{MNRAS}, 458, p. 1267-1288


\end{thebibliography}
\end{document}